\documentclass[preprint,proceedings]{rmaa}

\SetYear{2004}
\SetConfTitle{The Environment and Evolution of Binary and Multiple Stars}

\title{The Possible Belts for Extrasolar Planetary Systems}

\author{Ing-Guey Jiang\altaffilmark{1}, M. Duncan\altaffilmark{2}   
  and  D.N.C. Lin\altaffilmark{3}}

\altaffiltext{1}{National Central Univ., Taiwan,jiang@astro.ncu.edu.tw}
\altaffiltext{2}{Queen's University, Canada, duncan@astro.queeensu.ca}
\altaffiltext{3}{University of California, USA, lin@ucolick.org}

\shortauthor{Jiang, Duncan and Lin}
\shorttitle{Extrasolar Planetary Systems}

\fulladdresses{
\item Ing-Guey Jiang: Institute of Astronomy, National Central University,
 Taiwan (\email{jiang@astro.ncu.edu.tw}).
\item M. Duncan: Department of Physics, Queen's University, Kingston, ON
 K7L 3N6, Canada\\ (\email{duncan@astro.queeensu.ca}).
\item D.N.C. Lin: UCO/Lick Observatory, University of California, Santa Cruz,
 CA 95064, USA (\email{lin@ucolick.org}).}
\listofauthors{Ing-Guey Jiang, M. Duncan,  D.N.C. Lin}
\indexauthor{Jiang, I.-G.}
\indexauthor{Duncan, M.} 
\indexauthor{Lin, D.N.C.}

\addkeyword{Planets - extrasolar}

\abstract{More than 100 extrasolar planets have been discovered since 1990s.
Different from the solar system, these planets' orbital eccentricities
cover a huge range from $0$ to $0.7$. Incidently, the first Kuiper Belt Object
was discovered in 1992. Thus, an interesting and important question
will be whether extrasolar planetary systems could have structures 
like Kuiper Belt or asteroid belt. We investigate the stability of these 
planetary systems with different orbital eccentricities by the similar 
procedures in Rabl \& Dvorak (1988) and Holman \& Wiegert (1999).
We claim that most extrasolar planetary systems can have their own belts
at the outer regions. However, we find that the orbits with high--eccentricity
is very powerful in depletion of these populations.}

\resumen{Desde la d\'ecada de los 90 se han descubierto m\'as de 100 planetas extrasolares.
A diferencia del Sistema Solar, estos planetas tienen excentricidades en un
amplio intervalo, desde $0$ hasta $0.7$.  El primer objeto del Cintur\'on de
Kuiper se descubrió en 1992. Se plantea la cuestion de si los sistemas
planetarios
extrasolares podr\'\i an tener estructuras como el Cintur\'on de Kuiper o el de los
asteroides.  Investigamos la estabilidad de estos sistemas para distintas
excentricidades con los m\'etodos de  Rabl \& Dvorak (1988) y Holman \& Wiegert
(1999).  Sostenemos que la mayor parte de los sistemas planetarios extrasolares
pueden tener cinturones en las regiones externas.  No obstante, encontramos
que las orbitas de gran excentricidad son muy efectivas para destruir estas
estructuras.
}

\begin{document}
\maketitle 

\section{Introduction}

In the recent years, the number of discovered extra-solar planets is 
increasing quickly due to astronomer's observational effort and therefore 
the interest in dynamical study in this field has been renewed. 

These discovered planets with masses from 0.16 to 17 Jupiter masses ($M_J$) 
have semimajor axes from 0.04 AU to 4.5 AU and also a wide range
of eccentricities. Moreover, There is a mass-period correlation for discovered 
extra-solar planets, which gives the paucity of massive close-in 
planet. Jiang et al. (2003) claimed that although tidal interaction could
explain this paucity (P\"atzold \& Rauer 2002), the mass-period correlation 
might be weaker at the time when these planets were just formed. Gu et al. 
(2003)  and Sasselov (2003) also have done very interesting work on 
close-in planets. Therefore, some of these extrasolar planets' dynamical
properties are very different from the planets in the solar system.
  
Nevertheless, the similarity between extrasolar and solar planets do exist. 
For example, there is a new discovery about Jupiter-like orbit very recently,
i.e. a Jupiter-mass planet on a circular long-period orbit 
(semimajor axis $a$=3.65 AU) was detected.
 
On the other hand, some planetary systems were claimed to have discs of dust 
and they are regarded to be young analogues of the Kuiper Belt. For example,
Greaves et al. (1998) found a dust ring around a nearby star $e$ Eri 
and Jayawardhana et al. (2000) detected the dust in the 55 Cancri planetary 
system. Particularly, $\beta$ Pictoris planetary system has a warped disc 
and the influence of a planet might explain this warp (Augereau et al. 2001). 

Given the fact that many extrosolar planets' orbital eccentricities
are very big and some of them still could have analogues of the Kuiper Belt,
it would be interesting to investigate that what environments and conditions
could the belts exist for a planetary system. Following Rabl \& Dvorak 
(1988) and Holman \& Wiegert (1999), we use critical semimajor 
axis as a tool to explore the unstable zone where it would be difficult 
for a belt to exist  for a given planetary system.  

We will explain the basic model in Section 2. In Section 3, we study the 
cases of one planet. We discuss multiple planetary systems in Section 4 
and the effect of a companion star in a binary system in Section 5. We 
make conclusions and also discuss the possible implications in Section 6.
 
\section{The Model}

A direct force integration of the equation of motion is required for the 
computation of the orbital evolution of our systems.  We adopt a numerical 
scheme with Hermite block-step integration which has been developed by 
Sverre Aarseth (Markino \& Aarseth 1992, Aarseth, Lin \& Palmer 1993).

We consider a range of ratio ($\mu = M_p / (M_p + M_\ast)$) of masses
($M_p$ and $M_\ast$), where $M_p$ is the planetary mass and $M_\ast$ is the 
mass of the central star. We also consider a range of orbital eccentricity 
($e_p$). The semimajor axis of the (inner) planet is set to be unity
for systems with one (two) planets such that all other length scales are 
scaled with its physical value. We adopt $G (M_\ast + M_p) =1$ such that 
the planetary orbital period is $2 \pi$.

We mainly determine the inner and outer critical semimajor axis, i.e. 
the innermost and outermost semimajor axes at which the test particles 
both at $\theta=0^{\circ}, 90^{\circ}$ survive. 
The definition of survival here is that the distance between the test 
particle and the central star must be smaller than a critical value $R_d$ 
during a time $T_d$. The value of $R_d$ is arbitrarily set to be 3 times 
of planetary initial semimajor axis and we choose $T_d  = 2\pi \times 10^4$.
Therefore, more precisely, the inner critical semimajor axis is: within 
the region between the planet and central star, the outermost semimajor axis 
that a test particle can survive for $T_d$, and the outer critical semimajor 
axis is: out of the region between the planet and central star, the innermost 
semimajor axis that a test particle can survive for $T_d$.
  
Based on several test runs, we find that the value of $a_c$ does not
change significantly if $T_d$ is increased to $2\pi \times 10^6$.  That is,
planets which can survive for $2\pi \times 10^4$  can usually remain attached 
for a much longer timescale. Thus, we find critical semimajor axis $a_c$ to be 
a useful parameter to classify our results (Dvorak 2004).

\section{The Systems of One Planet}

We determine both the inner and outer critical semimajor axes for a system 
with one planet moving around the central star. The area between inner and 
outer critical semimajor axes can be regarded as ``unstable zone''. We get 
the width of unstable zone by subtracting the value of inner critical 
semimajor axis from the value of outer critical semimajor axis.

We determine these critical semimajor axes for different planetary mass.
We  also  consider different eccentricities of planet's orbits, which vary 
from $e = 0$ to $e = 0.8$. The results are in Tables 1a, 1b and 1c.

\linespread{.8}
\begin{table}[!t]
\begin{center}
\renewcommand{\arraystretch}{1.2}
\begin{tabular}{lcc}
\multicolumn{3}{c}{Table 1a}\\
\multicolumn{3}{c}{Critical Semimajor Axis When $\mu=0.005$ }\\
\toprule
        & inner & outer \\
\midrule
$e=0.0$ & 0.7  & 1.5 \\
$e=0.2$ & 0.5  & 1.9 \\	
$e=0.4$ & 0.3  & 2.1 \\
$e=0.6$ & 0.2  & 2.2 \\	
$e=0.8$ & 0.1  & 2.5 \\
\bottomrule
\end{tabular}
\vspace*{\baselineskip}
\renewcommand{\arraystretch}{1.2}
\begin{tabular}{lcc}
\multicolumn{3}{c}{Table 1b}\\
\multicolumn{3}{c}{Critical Semimajor Axis When $\mu=0.001$ }\\
\toprule
        & inner & outer \\
\midrule
$e=0.0$ & 0.8  & 1.3 \\
$e=0.2$ & 0.6  & 1.6 \\	
$e=0.4$ & 0.4  & 1.8 \\
$e=0.6$ & 0.2  & 2.1 \\	
$e=0.8$ & 0.1  & 2.2 \\
\bottomrule
\end{tabular}
\vspace*{\baselineskip}
\renewcommand{\arraystretch}{1.2}
\begin{tabular}{lcc}
\multicolumn{3}{c}{Table 1c}\\
\multicolumn{3}{c}{Critical Semimajor Axis When $\mu=0.0001$ }\\
\toprule
        & inner & outer \\
\midrule
$e=0.0$ & none & none\\
$e=0.2$ & 0.7  & 1.3 \\	
$e=0.4$ & 0.5  & 1.6 \\
$e=0.6$ & 0.3  & 1.8 \\	
$e=0.8$ & 0.1  & 1.9 \\
\bottomrule
\end{tabular}
\end{center}

\vspace*{-.75cm}
\end{table}
\linespread{1}

If the mass of the central star is assumed to be 1 $M_\odot$, the  
planet has about 5 $M_J$ for the results in Table 1a
and has about 1 $M_J$ for the results in Table 1b.
From Tables 1a and 1b, we find
that the results are quite similar for these two cases.
Approximately, the inner critical semimajor 
axis is about $3/4$  and the outer 
critical semimajor axis is about ${3/2}$ when the eccentricity $e=0$. 
After we increase the eccentricity, 
the inner critical semimajor axis become about $(1-e)3/4$ and
the outer critical semimajor axis become about $(1+e)3/2$.
This is reasonable because the peri-centre is at $(1-e)a$ and the 
apo-centre is at $(1+e)a$ where $a$ is the semimajor axis of the planet
and thus the planet's orbit covers a larger radial range,
the critical semimajor axis should change correspondingly.  

However, from the results in Table 1c, 
when the mass of the planet is much less (one order less)
than $M_J$, the planet depletes nothing and thus both 
the inner and outer critical semimajor axes do not exist in the case
of zero eccentricity. Interestingly, when we increase the eccentricity,
the effect of eccentricity gradually dominates and critical semimajor
axes can become similar order as the ones in Table 1a and 1b 
even the mass of the planet is much less.

\section{The Systems of Two Planets}

Interestingly, there are two belts of small bodies in the solar system and 
these two are located in very different environments: the asteroid belt
is between two planets and the Kuiper Belt is located at the outer part of 
the planetary disc. Therefore, it will be important to study multiple 
planetary systems and determine the  physical locations where we can possibly
have stable belts.
 
To simplify the model and as a first step, we choose the case of two planets 
and both with mass about $M_J$, i.e. $\mu=0.001$. The inner planet will be 
planet 1 and the outer planet will be planet 2 hereafter. The stability of 
this system depends on their separation and also orbital eccentricities. To 
reduce the parameters, we always set the initial eccentricity of planet 2 to be
zero but study the effect of different initial eccentricity of planet 1 only. 

These two planets are in fact interacting to each other. When the initial 
eccentricity of planet 1 is small, the interaction is weaker and planet 2 
stays to move on a nearly circular orbit.  When the initial eccentricity of 
planet 1 is bigger, the interaction becomes much stronger and planet 2 
gradually increases the eccentricity of its orbit in our simulations.
 
We checked the critical semimajor axis of planet 2 for different 
eccentricities of planet 1 and we found that when planet 1 is initially 
located at $r=1$, planet 2 should be about $r=3$ to make whole system 
stable during $10^4$ rotation periods of planet 1.

Therefore, we set the semimajor axis of planet 1 to be unity, 
the semimajor axis of planet 2 to be 3 and both at $\theta=0^\circ$ initially.
We then begin to put test particles to determine the inner and outer
critical semimajor axes for both planets. Tables 2a and 2b are the results.

When the eccentricity is 0 or 0.2, we found that there could be an asteroid 
belt-like population between these two planets. The system is quite stable and
thus the results in last section, i.e. Table 1b, gives us good hints for the 
size of unstable zone around planet 1 though the unstable zone does expand 
a bit for this system with two planets. However, when the eccentricity is 
larger, there is no stable zone between these two planets and the most 
possible location to have a belt is out of the outer critical semimajor 
axis of planet 2. This result tells us that if the eccentricities of the 
planets in the solar system is not between 0 and 0.2, but much larger, it is 
unlikely that there would be an asteroid belt. 

\linespread{.8}
\begin{table}
\begin{center}
\renewcommand{\arraystretch}{1.2}
\begin{tabular}{lcc}
\multicolumn{3}{c}{Table 2a}\\ 
\multicolumn{3}{c}{Critical Semimajor Axis of Planet 1}\\
\toprule
        & inner  & outer\\
\midrule
$e=0.0$ & 0.7  & 1.3    \\
$e=0.2$ & 0.6  & 1.7    \\	
$e=0.4$ & 0.3  & none   \\
$e=0.6$ & 0.2  & none   \\  	
$e=0.8$ & 0.1  & none  \\ 
\bottomrule
\end{tabular}
\renewcommand{\arraystretch}{1.2}
\begin{tabular}{lcc}
\multicolumn{3}{c}{Table 2b}\\
\multicolumn{3}{c}{Critical Semimajor Axis of Planet 2}\\
\toprule
        & inner  & outer \\
\midrule
$e=0.0$   & 2.3   & 3.9 \\
$e=0.2$  & 2.3   & 3.9 \\	
$e=0.4$  & none & 3.9 \\
$e=0.6$  & none & 3.9  \\  	
$e=0.8$  & none & 7.5  \\ 
\bottomrule
\end{tabular}
\end{center}
\vspace*{-.5\baselineskip}
\end{table}
\linespread{1}
\section{The Effect of a Companion Star}

Some of the host stars of the discovered planetary systems are indeed members 
of binary systems, for example, 16 Cyg B, 55 $\rho^1$ Cnc, $\tau$ Boo.

It will be interesting to see the effect of a secondary companion star  on the 
planetary system in which a planet moves around the binary primary. Thus, 
assuming an equal mass binary, we determine the critical semimajor axis of 
binary secondary for both the cases that the eccentricity of the binary 
$e_{\rm b}$ is 0.2 and 0.6. We assume the planet has mass about 1 $M_J$, 
the initial eccentricity ranges from 0 to 0.8 with respect to the binary 
primary. Both the binary secondary and the planet begin from 
$\theta=0^\circ$. Tables 3a and 3b are our results and we found 
that most of the discovered planets are stable since their binary separations 
are much bigger than the critical semimajor axes.
 
On the other hand, the binary might affect the extension of possible belt 
populations. To further investigate this point, we now use two test particles 
(at $\theta=0^\circ$ and $90^\circ$) to determine the critical semimajor axis 
of binary secondary. We find that the critical semimajor axis of the system 
becomes bigger in order to make the test particles survive. For the case 
of $e_{\rm b}=0.2$, the critical semimajor axis $a_{\rm b}$ is about 20.
For the case of $e_{\rm b}=0.6$, $a_{\rm b}$ is about 36. We find that 
this result does not depend on the details of other parameters such as the 
initial eccentricity of planet or semimajor axis of test particles.
Because the critical semimajor axis becomes much bigger, the presence of a 
binary secondary might affect the extension or even existence of possible 
Kuiper Belt populations.

\linespread{.7}
\begin{table}
\begin{center}
\begin{tabular}{lc}
\multicolumn{2}{c}{Table 3a}\\
\multicolumn{2}{c}{Critical Semimajor Axis When $e_{\rm b}=0.2$ }\\
\toprule
        & $a_{\rm b}$  \\
\midrule
$e=0.0$ & 4.1 \\
$e=0.2$ & 4.5 \\	
$e=0.4$ & 4.9 \\
$e=0.6$ & 5.2 \\	
$e=0.8$ & 5.4 \\
\bottomrule																											
\end{tabular}
\vspace*{\baselineskip}
\renewcommand{\arraystretch}{1.2}
\begin{tabular}{lc}
\multicolumn{2}{c}{Table 3b}\\
\multicolumn{2}{c}{Critical Semimajor Axis When $e_{\rm b}=0.6$ }\\
\toprule
        & $a_{\rm b}$  \\
\midrule
$e=0.0$ & 9.7 \\
$e=0.2$ & 10.1 \\	
$e=0.4$ & 10.6 \\
$e=0.6$ & 11.1 \\	
$e=0.8$ & 11.5 \\
\bottomrule
\end{tabular}
\end{center}
\vspace*{-\baselineskip}
\end{table} 
\linespread{1}

\section{Conclusions and Implications}
We have studied the possible conditions that a belt could be stable and 
thus exist for assumed planetary systems. Because we explore different 
eccentricities for the given planetary systems, our results shall be able 
to apply to discovered extrasolar planetary systems. In addition to those 
systems with one planet, systems with two planets are studied and the effect 
of a companion star is also investigated.

We find that highly eccentric orbits are very powerful in depletion of
the belt-like population such as asteroid belt and the companion star
might restrict the extension of such populations.

On the other hand, as emphasis by Yeh \& Jiang (2001), the planet should 
dynamically couple with the belt over the evolutionary history. That is, 
the planet's mass and orbital properties would determine the existence and 
affect the position of the belt, but if the belt is massive enough, it will in
turn influence the planet, too.  This is particularly important during the 
early stage of planetary formation since the circumstellar belt is more 
massive at that time. For example, Jiang \& Ip (2001) mentioned that 
the interaction with the belt could bring the planetary system of upsilon 
Andromedae to the current orbital configuration.

Moreover, according to Jiang \& Yeh (2003a), the probability that the planet 
moves stablely around the outer edge is much smaller than near the inner edge.
This conclusion is consistent with the principal result in Jiang \& Yeh (2003b).

What could we learn for the solar system from their theoretical result ? 
From the observational picture of the asteroid belt, we know that: (a) the 
outer edge looks more diffuse and (b) Mars is moving stably close to the 
inner edge of the asteroid belt but
\adjustfinalcols
 Jupiter is quite far from the outer edge. 
One possible explanation is that since Jupiter is much more massive and thus 
those planetesimals close to Jupiter would have been scattered away during the
formation of the solar~system.
If we apply the  model of Jiang \& Yeh (2003a) on the asteroid belt and the 
point mass which represents the planet in their model can also represent 
larger asteroids, their theoretical result provides another choice to 
explain both (a) and (b).

It is  known that there is another belt in the solar system, the Kuiper Belt,  
after the first object was detected (Jewitt \& Luu 1993).
 Allen, Bernstein \& Malhotra (2001) did a survey and 
found that they could not find any Kuiper 
Belt Objects (KBOs) larger than 160 km in diameter 
beyond 50 AU in the outer solar system. 
If we apply the model of Jiang \& Yeh (2003a) on this problem and the point 
mass which represents the planet in their model now represents larger KBOs 
moving within the Kuiper Belt, we find that their  theoretical results
provide a natural mechanism to do this orbit rearrangement:
larger KBOs might have been moving towards the inner edge of the belt 
due to the influence from the belt.  
\vspace*{-.05cm}

\end{document}